\documentstyle[12pt]{article}
\vspace{1.8cm} \oddsidemargin 0pt \evensidemargin 0pt  \topmargin
-1.5cm \footheight 26pt \footskip 10mm

\begin{document}
\baselineskip 20pt
\begin{center}
\baselineskip=24pt {\Large  Finite Temperature Schr\"{o}dinger
Equation}

\vspace{1cm} {Xiang-Yao Wu$^{a}$ \footnote{E-mail:wuxy2066@163.com
}, Bai-Jun Zhang$^{a}$, Xiao-Jing Liu$^{a}$\\ Nuo Ba$^{a}$,
Yi-Heng Wu$^{a}$, Qing-Cai Wang$^{a}$, Yan Wang$^{a}$} \vskip 10pt
\noindent{\footnotesize a. \textit{Institute of Physics, Jilin
Normal University, Siping 136000, China}}

\end{center}
\date{}
\renewcommand{\thesection}{Sec. \Roman{section}} \topmargin 10pt
\renewcommand{\thesubsection}{ \arabic{subsection}} \topmargin 10pt
{\vskip 5mm
\begin {minipage}{140mm}
\centerline {\bf Abstract} \vskip 8pt
\par
\indent\\

\hspace{0.3in} We know Schr\"{o}dinger equation describes the
dynamics of quantum systems, which don't include temperature. In
this paper, we propose finite temperature Schr\"{o}dinger
equation, which can describe the quantum systems in an arbitrary
temperature. When the temperature $T=0$, it become Shr\"{o}dinger
equation. \\
\vskip 5pt

PACS numbers: 05.45.-a; 11.10.Wx\\
Keywords: Schr\"{o}dinger equation; Finite temperature

\end {minipage}

\newpage
\section * {1. Introduction }

\hspace{0.3in}

High-temperature and dense matter of elementary particles appears
in several areas of physics. The first and most familiar example
is the Universe at the early stages of its expansion. The Big Bang
theory states that the Universe was hot and dense in the past
[1,2], with a temperature ranging from a few $eV$ up to the Planck
scale. Another place, where matter exists in extreme conditions of
high densities, can be created in the laboratory. Namely, in
heavy-ion collisions extended dense fireballs of nuclear matter
are created, with the energy density exceeding the QCD scale [3].

From a practical point of view, there is quite a wide area of
applications of finite-temperature field theory to cosmology and
laboratory experiments. The high-temperature phase transitions,
typical for grand unified theories (GUTs), may be important for
cosmological inflation and primordial density fluctuations.
Topological defects can naturally arise at the phase transitions
and influence the properties of the Universe we observe today. The
first-order electroweak (EW) phase transition is a crucial element
for electroweak baryogenesis; it may also play a role in the
formation of the magnetic fields observed in the Universe. The QCD
phase transition and properties of the quark-gluon plasma are
essential for the understanding of the physics of heavy-ion
collisions. The QCD phase transition in cosmology may influence
the spectrum of the density fluctuations relevant to structure
formation.

The construction of an effective action for quantum
electrodynamics (QED) in the presence of various external
conditions has been a challenge since the early days of the
theory. The study of generalizations of the Heisenberg-Euler
Lagrangian that include finite temperature effects has been
initiated by Dittrich [4], who considered the case of a constant
external magnetic field at finite temperature using the imaginary
time formalism. An extension of this work to the case of arbitrary
constant electromagnetic fields turned out to be qualitatively
more substantial than naively expected. Employing the real-time
formalism, this situation was investigated by Cox, Hellman and
Yildiz [5], and Loewe and Rojas [6]. A more comprehensive study of
the problem has been performed by Elmfors and Skagerstam [7], who
corrected the preceding findings and additionally introduced a
chemical potential. An attempt employing the imaginary-time
formalism was made by Ganguly, Kaw and Parikh [8] for the case of
an external electric field. Recently, the finite-temperature
effective action for electromagnetic fields was studied by
Shovkovy [9] in the worldline approach, where finite temperature
is also introduced via an imaginary-time formalism. In this paper,
based on the first law of thermodynamics, we define the
microscopic entropy of single particle and multi-particle, and
give the finite temperature Schr\"{o}dinger equations for the
single particle and multi-particle, respectively.

\section * {2. The concept of microscopic and macroscopic entropy}

\hspace{0.3in} In thermodynamics, the infinitesimal entropy change
$dS$ of a system is defined by
\begin{equation}
dS=\frac{\delta Q}{T},
\end{equation}
where $\delta Q$ is a transfer of heat between the composite
system and an external reservoir at the temperature $T$. In a
finite change of state from $i$ to $j$, the entropy change is
$S_{j}-S_{i}$,
\begin{equation}
S_{j}-S_{i}=\int^{j}_{i}\frac{\delta Q}{T},
\end{equation}
The entropy of a system is a function of the thermodynamic
coordinates whose change is equal to the integral of $\frac{\delta
Q}{T}$ between the terminal states, integrated along any
reversible path connecting the two states.

By integrating (1) around a reversible cycle, so that the initial
and final entropies are the same. For a reversible cycle, we get
\begin{equation}
\oint\frac{\delta Q}{T}=0.
\end{equation}
In quantum statistics, the entropy $S$ is defined by
\begin{equation}
S=-k_{B}Tr(\rho \ln\rho),
\end{equation}
where $\rho=|\psi\rangle\langle\psi|$ is the density matrix,
$k_{B}$ is the Boltzmann constant.

In classical statistics, the entropy is defined by the Boltzmann
\begin{equation}
S=k_{B}\ln W=-k_{B}\ln\rho,
\end{equation}
where $W$ is the total number of the possible microscopic states,
and $\rho=\frac{1}{W}$ is probability of every state.

From (5), we know that the macroscopic entropy is from the state
distribution of a large number of particles. According to the
viewpoint of macroscopic entropy, as the single particle hasn't
state distribution, the single particle hasn't entropy.

In the viewpoint of quantum mechanics, a microscopic particle has
wave-particle duality, and the wave nature is described by wave
function. Obviously, the wave functions have the nature of
distribution. So, a single particle has microscopic entropy, and
we define its microscopic entropy $S$ as

\begin{equation}
S=-k_{w}\ln|\psi(\vec{r},t)|^{2},
\end{equation}
where $k_{w}$ is a constant, which will be confirmed by
experiment. The total microscopic entropy $S$ is a functional of
wave function $\psi$. It can be written as
\begin{equation}
S[\psi(\vec{r},t)]=-k_{w}\ln|\psi(\vec{r},t)|^{2}.
\end{equation}
For the multi-particle system, the system total entropy is
\begin{equation}
S=-k_{w}\ln|\psi(\vec{r}_{1}, \vec{r}_{2},\cdots \vec{r}_{N},
t)|^{2}.
\end{equation}
where $\psi(\vec{r}_{1}, \vec{r}_{2},\cdots \vec{r}_{N}, t)$ is
the total wave function of multi-particle system.

\section * {3. Finite temperature Schr\"{o}dinger Equation}
Based on the first law of thermodynamics, there is
\begin{equation}
dU=dW+dQ,
\end{equation}
where $dU$ is the infinitesimal change of system internal energy,
$dW$ is the infinitesimal work doing by surrounding, and $dQ$ is
the infinitesimal heat absorbed from surrounding. In this paper,
we should study the quantum thermodynamics property of microscopic
particles, i.e., finite temperature quantum theory of N-particle.
When $dW=0$, the (9) becomes
\begin{equation}
dU-dQ=0.
\end{equation}
For the point particle system of N-particle, the internal energy
$U$ is
\begin{equation}
U=\sum_{i=1}^{N}(T_{i}+V_{i})+\sum_{i<j}^{N}V_{ij},
\end{equation}
where $T_{i}$ and $V_{i}$ are the i-th particle's kinetic energy
and potential energy, and $V_{ij}$ is the interaction energy of
i-th and j-th particle, the internal energy infinitesimal change
is
\begin{equation}
dU=d(\sum_{i=1}^{N}(T_{i}+V_{i})+\sum_{i<j}^{N}V_{ij}),
\end{equation}
since
\begin{equation}
dQ=\sum_{i=1}^{N}dQ_{i}=\sum_{i=1}^{N}TdS_{i}=d(\sum_{i=1}^{N}TS_{i}),
\end{equation}
where $dQ_{i}$ is the i-th particle absorbed heat, $dS_{i}$ is the
i-th entropy change. $T$ is the system temperature. \\
By substituting (12), (13) into (10), we have
\begin{equation}
d(\sum_{i=1}^{N}(T_{i}+V_{i})+\sum_{i<j}^{N}V_{ij}-T\sum_{i=1}^{N}S_{i})=0.
\end{equation}
i.e.,
\begin{equation}
\sum_{i=1}^{N}(T_{i}+V_{i})+\sum_{i<j}^{N}V_{ij}-TS=E,
\end{equation}
where $S=\sum_{i=1}^{N}S_{i}$ is system total entropy, $E$ is a
constant, which can be defined as the total energy of system. \\
By canonical quantization
\begin{equation}
E\rightarrow i\hbar \frac{\partial}{\partial t}, \hspace{0.3in}
T_{i}=-\frac{\hbar^{2}}{2m_{i}}\nabla_{i}^{2},
\end{equation}
and
\begin{equation}
S=-k_{w}\ln|\psi(\vec{r}_{1}, \vec{r}_{2},\cdots \vec{r}_{N},
t)|^{2}.
\end{equation}
The (15) becomes quantum wave equation
\begin{eqnarray}
&&i\hbar \frac{\partial}{\partial t}\psi(\vec{r}_{1},
\vec{r}_{2},\cdots \vec{r}_{N},
t)=[\sum_{i=1}^{N}(-\frac{\hbar^{2}}{2m_{i}}\nabla_{i}^{2}+V_{i})+\sum_{i<j}^{N}V_{ij}
\nonumber\\&&+k_{w}T\ln|\psi(\vec{r}_{1}, \vec{r}_{2},\cdots
\vec{r}_{N}, t)|^{2}]\times\psi(\vec{r}_{1}, \vec{r}_{2},\cdots
\vec{r}_{N}, t).
\end{eqnarray}
The (18) is finite temperature Schr\"{o}dinger equation of
multi-particle. When $T=0$, the (18) becomes Schrodinger
equation of multi-particle system.\\
The partial differential (18) can be solved by the method of
separation of variable. By writing
\begin{equation}
\psi(\vec{r}_{1}, \vec{r}_{2}\cdots \vec{r}_{N},
t)=\psi(\vec{r}_{1}, \vec{r}_{2}\cdots \vec{r}_{N})g(t),
\end{equation}
submitting (19) into (18), we have

\begin{equation}
i\hbar\frac{dg(t)}{dt}-k_{w}T\ln g^{2}(t)  \cdot g(t)=E g(t),
\end{equation}
and

\begin{eqnarray}
&&\sum^{N}_{i=1}(-\frac{\hbar^{2}}{2m_{i}}\nabla^{2}\psi(\vec{r}_{1},
\vec{r}_{2}\cdots \vec{r}_{N})+V(r_{1}, r_{2}\cdots
r_{N})\psi(\vec{r}_{1}, \vec{r}_{2}\cdots \vec{r}_{N})
 \nonumber\\&&+\sum^{N}_{i<j}V_{ij}\psi(\vec{r}_{1},
\vec{r}_{2}\cdots \vec{r}_{N})+k_{w}T(\ln|\psi(\vec{r}_{1},
\vec{r}_{2}\cdots \vec{r}_{N})|^{2} )\psi(\vec{r}_{1},
\vec{r}_{2}\cdots \vec{r}_{N})\nonumber\\&&=E\psi(\vec{r}_{1},
\vec{r}_{2}\cdots \vec{r}_{N}).
\end{eqnarray}
Equation (21) is time-independent finite temperature
Schr\"{o}dinger equation of multi-particle system.

For the single particle, the (15) becomes
\begin{equation}
\frac{p^{2}}{2m}+V-TS=E,
\end{equation}
where $\frac{p^{2}}{2m}$, $V$ and $E$ are the single particle
kinetic energy, potential energy and total energy, $S$ is the
single particle total microscopic entropy and $T$ is the single
particle's surrounding temperature. For a classical particle, its
entropy is zero, and its total energy $E=\frac{p^{2}}{2m}+V$,
i.e., its total energy is mechanical energy. For a microscopic
particle, due to the wave nature, it has microscopic entropy. The
microscopic particle total energy $E$ is the sum of its mechanical
energy $\frac{p^{2}}{2m}+V$ and $-TS$. We define $-TS$ as heat potential energy.\\
By canonical quantization
\begin{equation}
E\rightarrow i\hbar \frac{\partial}{\partial t}, \hspace{0.3in}
T=-\frac{\hbar^{2}}{2m}\nabla^{2},
\end{equation}
and
\begin{equation}
S=-k_{w}\ln|\psi(\vec{r},t)|^{2},
\end{equation}
the (22) becomes quantum wave equation

\begin{equation}
 i\hbar \frac{\partial}{\partial t}\psi(\vec{r},t)=-\frac{\hbar^{2}}{2m}\nabla^{2}\psi(\vec{r},t)
 +V\psi(\vec{r},t)+k_{w}T(\ln|\psi(\vec{r},t)|^{2})\psi(\vec{r},t).
\end{equation}
The (25) is the finite temperature Schrodinger equation of
single-particle. When $T=0$, it becomes the single particle
Schrodinger equation.

The partial differential (25) can be solved by the method of
separation of variable. By writing
\begin{equation}
\psi(\vec{r}, t)=\psi(\vec{r})g(t),
\end{equation}
submitting (26) into (25), we have
\begin{equation}
i\hbar\frac{dg(t)}{dt}-k_{w}T\ln g^{2}(t) \cdot g(t)=E g(t),
\end{equation}
and

\begin{eqnarray}
-\frac{\hbar^{2}}{2m}\nabla^{2}\psi(\vec{r})+V(r)\psi(\vec{r})+k_{w}T(\ln|\psi(\vec{r})|^{2}
)\psi(\vec{r})=E\psi(\vec{r}).
\end{eqnarray}
Equation (28) is time-independent finite temperature
Schr\"{o}dinger equation of single-particle.

\section * {4. Conclusion}
\hspace{0.3in}In this paper, we define microscopic entropy of
single particle and multi-particle, and give the finite
temperature Schr\"{o}dinger equation of a single and
multi-particle. With these equations, we can study quantum systems
in an arbitrary temperature, such as superconductivity mechanism,
Bose-Einstein condensates and so on.

\end{document}